# On the Design and Analysis of a Biometric Authentication System using Keystroke Dynamics

Robert Cockell and Basel Halak
*Electronics and Computer Science*
*University of Southampton*
Southampton, England
Emails {basel.halak@soton.ac.uk, rc8g15@soton.ac.uk}

*Abstract*—this paper proposes a portable hardware token for user's authentication; it is based on the use of keystroke dynamics to verify users biometrically. The proposed approach allows for a multifactor authentication scheme, in which a user cannot be granted access unless they provide a correct password on a hardware token and their biometric signature. The latter is extracted while the user is typing their password. This paper explains the design rationale of the proposed system and provides a comprehensive insight in the development of a hardware prototype of the same. The paper also presents a feasibility study that included a systematic analysis based on training data obtained from 32 users. Our results show that dynamic keystroke can be employed to construct a cost-efficient solution for biometric user authentication with an average error rate of 4.5%

I. INTRODUCTION

To use a computer securely, a user must prove their identity to the machine in question. The vast majority of systems use a password-based (or passcode-based) authentication mechanism. Such a system, however, has a glaring hole in its security – the computer grants access if the password is correct, in other words, it only verifies the correctness of the presented password, and crucially it does not check the identity of the intended user. Therefore, if a malicious third-party guesses or somehow deduces the user's password, the entire security of the system is compromised. There are many examples in the literature that exploit this vulnerability to break the security of password-based systems, example of these include the brute-force attack reported in in [1].

An alternative method of authentication involves having a physical item that the user must produce to be given access, for example a key card. A more secure version of this is based on the use of hardware tokens, to identify the user and allow access. These hardware token require two or even three factors for authentication: for example, a card reader for online banking may require the presence of the reader, the presence of the bankcard and the correct pin number to log in. There has been plenty of previous research into systems involving a hardware token and using one securely as part of a larger system [2], [3-9]. This increases the strength of the authentication by needing multiple factors to be passed before access is allowed, however none of these factors require the intended user to be the one logging in. In all cases, only the presence of data or hardware is being verified: the identity person providing them is not checked, therefore if the hardware token is lost, stolen or forged, the computer's security collapses.

To solve this problem, the computer must check for the user directly. Rather than detecting the presence something the user knows, like a passcode, or something the user has, like a key card, it must check for something that the user is. Authenticating the user biometrically makes for some of the most secure systems available. Examples include a fingerprint scanner in [10], such a system is reported to achieve false rejection rate (FRR) less than 4.12% when restricting the false acceptance rate (FAR) to 0.01%. Another biometric-based identification scheme is the retina scanners in [11], which achieves a 0% FAR and 1.85% FRR. These systems recognise the user through their body, removing the risk of stolen data or keys, as well as the risk of a password being guessed. However, these systems are prohibitively expensive requiring complex hardware to make the biometric measurement and to carry out necessary data processing to identify the user. This work proposes an alternative biometric method for user authentication based on the use of keystroke dynamics. When people type, on a keyboard or on a number pad, they will do so with differing speeds, differing press and release times, and different pressures. What's more, a mixture of various factors affects these differences – hand size, muscle state, how one was taught to type and so on – and these factors are characteristic to each individual. Keystrokes can show much data about the user, one study even managing to predict user's emotional state when typing with around 85% accuracy [12].

The contributions of this work are as follows:

1) The developmnet of a biometric-based multifactor user authentication scheme, which expolits keystrock dynamics
2) The design and implmentation of a hardware prototype of the proposed system
3) The evaluation of the usability of the proposed system through a feasbility study, which are based on training data from 32 users

Our results show that keystroke dynamics can be used to build a cost-effective biometric based authentication scheme with an average error rate of 4.5%.

The remainder of this paper is organised as follows. Section 2 discussed the related works. Section explains the principles of the proposed approach. Section 4 discusses in depth the design and implementation of a hardware prototype of the system. Section 5 explains the technique employed to evaluate the usability of our solution. Finally, conclusions are drawn in section 6.

II. RELATED WORK

Keystroke dynamics as a field has seen a fair amount of research [13], with various different algorithms extracting various aspects of typing on various different platforms, including mobile phones [14]. The methodology behind keystroke dynamics follows much the same principle as the study of handwriting, but modern technology allows for multiple different ways of recognising the user. The most obvious way of extracting data on the user from keypresses is to measure the hold time, being how long the button is pressed for, and the space time, how long between keypresses. Results using this as a feature vary from a FAR of 0.5% and an FRR rate of 3.1% when only considering the space time, to equal error rates of between 4% and 12% depending on the algorithm [15]. When also considering the interval between the same point on successive keystrokes, a 1.45% FRR and a 1.89% FAR has been obtained [16].

Also worthy of note is systems that use continuous authentication in keystroke dynamics. These systems commonly also use the hold and space times, but rather than authenticating the user over the course of a password they will passively authenticate the user as they type. The aim is to secure a system against the user changing after the login process has ended by constantly keeping track of the user's keystroke dynamics, with result. While continuous authentication is not the focus of this paper, the methods and algorithms employed are similar to login authentication, for example Darabseh and Namin found that hold duration was a more effective identifying characteristic than latency time [17]. They produced an accuracy of around 80% when using support vector machines and k-nearest neighbour methods, however neural nets have produced equal error rates of around 2% [18].

To produce greater accuracy, other characteristics of the keypress can be extracted apart from hold and space times. When extracting keystroke sounds for continuous authentication, Roth et al. achieved an equal error rate of around 11% [19]. Another study using the sound of keystrokes over a sentence reported an FAR of 11% and an FRR of 12% [20]. While these results are interesting, recording the pressure or displacement of keypresses has achieved better results [13]. As noted in [21], research into this area is limited by the difficulty of creating a pressure sensing keyboard. Nevertheless, there have been a few studies into this area. One principle drawn from past studies is the idea of extracting attributes from the keypresses to have qualities to compare. Examples of how characteristics of the keypress can be used in this way are in [22] (which also measured vibration), and [23], with best results 1.67% FAR at 0% FRR for [23] and 0.6% error rate for [22], which was an extensive study using neural methods.

This paper differs from most papers using analogue keystroke dynamics because the system is on a keypad, not a full keyboard. Logically speaking, authenticating on a keypad would be harder, because there are fewer unique keys to compare against, and generally fewer fingers used when keying in passcodes. Possible implementations have been suggested [24], but error rates for previous systems have been relatively high, with an accuracy of only 40% [25], and an equal error rate of 10% for [26]. A comparison between this work and related ones is shown in table 1.

TABLE I. TABLE OF RELATED STUDIES

| Study | Type of study | Error rates |
| --- | --- | --- |
| Bleha et al. [27] | Keyboard timing data | FAR 0.5%, FRR 3.1% |
| Anusas-amornkul and Wangsuk [15] | Keyboard timing data | ERR 4-12% |
| Araujo et al. [16] | Keyboard timing data | FAR 1.89%, FRR 1.45% |
| Darabseh and Namin [17] | Continuous authentication | Around 20% error rate |
| Ahmed and Traore [18] | Continuous authentication | Around 2% error rate |
| Roth et al. [19] | Keystroke sounds | Around 11% error rate |
| Zhou et al. [20] | Keystroke sounds | FAR 11%, FRR 12% |
| Sulong et al. [23] | Keyboard pressure/displacement | FAR 1.67%, FRR 0% |
| Sulavko et al. [22] | Keyboard pressure/displacement | 0.6% error rate |
| Loh et al. [25] | Number pad pressure/displacement | 60% error rate |
| Grabham and White [26] | Number pad pressure/displacement | EER 10% |
| This study | Number pad pressure/displacement | 4.5% average error rate |

## III. PROPOSED SYSTEM ARCHITECTURE

The proposed system is envisioned to be a self-contained portable hardware token. It consists of a number pad connected to an embedded chip. The latter is mainly responsible for capturing the keystroke profile associated with each user, with this information stored in the memory for a given password/user. The system allows users to change their password and allows for multiple user entries.

Each time the user logs on, they will simply enter the password onto the device (logging on here is used as an example, any other use of verification would happen the same way, such as confirming a purchase). The press data from the keys is compared to the profile stored in memory, and if the data matches within a suitable threshold, the user is given access. If an unintended user attempts to log on, even with the correct password, their data will not match that stored in memory, and they will not be allowed access. This system already requires multiple factors for authentication (Password, number pad and matching press data), but could also require another token, such as a card or a key, for even greater security. The beauty of this method is that it will extract biometric signatures of users from while they type their password, therefore it does not require complicated circuitry associated with retina or fingerprint scanners, nor will it need the additional user input such systems require. To log onto a system a user only needs to enter a password onto the device, and merely doing so can be used to authenticate them. The architecture of the proposed system is divided into two main parts, hardware and software. The hardware part is responsible capturing the keystroke profile from the user's input and translating these analogue measurements into digital information. The software part has two functionalities; first it is responsible for extracting user's signature, which are stored in on-chip memory, and second it is responsible for authenticating users at a later stage. A generic system architecture is shown in figure 1. The operation principles are as follows. The user input is captured using motion sensors connected to a key pad or a key board (not shown in this figure). Next, the output of the sensing stage is fed into an ADC, which generates the digital data to be processed at a later stage. The comparator block provides a digital signal for each button press, with this signal used to indicate to the processor that there is data on the bus to be analysed/processed. This removes the need for the processor to be constantly active, which allows for a more energy efficient implementation. The digital signals are also used to make sure the entered password is correct. Finally, the processor samples the data from the bus and generates a profile for each user's password, which is subsequently stored in the memory block.

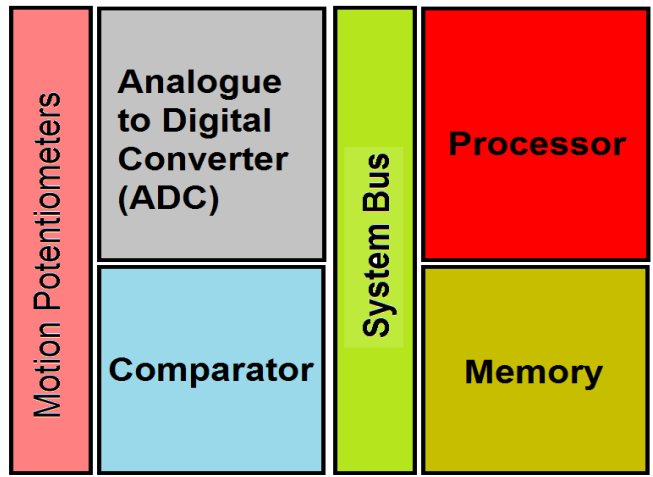

Fig. 1: A system block diagram

The essence of the proposed approach relies on the assumption that it is possible to distinguish between users based on their typing speed/style. To illustrate the feasibility of this technique, we carried a simple experiment wherein we measured the length of a key pad button press for ten users while typing a password. Fig. 2 shows how the results. It can be clearly seen that each user's presses are grouped around a certain range of lengths, and this differs for each user. Although there are some similarities, for example between users 5 and 7, this is just one characteristic for one button in a passcode. When considering all of the different factors it should be possible to verify users accurately. More details on this can be is in section (*V.A*).

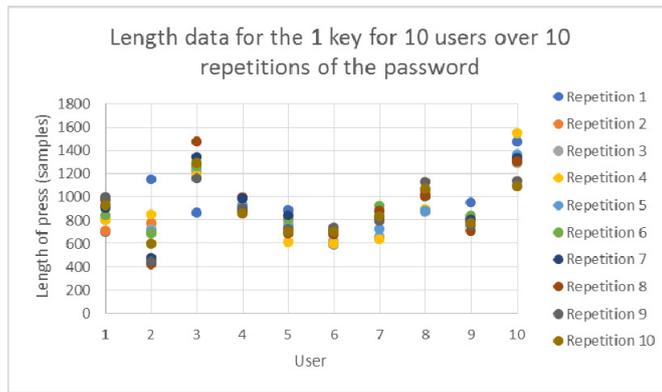

Fig. 2: A graph showing the clustering effect of the length of press of different users when typing the same password.

## IV. IMPLEMENTATION

This section explains the design rationale of a prototype of the proposed system.

### A. Key pad

In this case study, we choose to implement a key pad that has ten buttons. This seems a reasonable trade-off between the usability of the system and its size. In principle, the more buttons there are, the longer and more secure the passwords, the larger the number of possible passwords and the larger the size of the hardware token.

The buttons on the pad should also have reasonable stiffness and travel, so that the range of travel is roughly equal to the range of a normal button press. In addition, the keys should return as fast as possible after the press, ideally moving up with the finger that pressed the key, so the data recorded on the upstroke shows how the button was released, rather than how the button is returned to the resting position. The number pad should also contains analogue sensing elements; previous studies have used piezoelectric sensors [24], or force sensitive resistors [22, 25] to measure the pressure of a button press. In this case study, we chose linear motion potentiometers mostly chosen for the ease of creating buttons.

To allow for easy mounting of the analogue components the case for the number pad was 3D printed. Linear motion potentiometers act the same way as normal potentiometers (the resistance between a 'wiper' contact and two end contacts varies with position); the only difference being the position here is the linear position of a shaft, rather than radial position. The design of the case had slots that the potentiometers fit into, with the wires accessible from the bottom and the shafts free to move. The key caps were also 3D printed, so the diameter of the screw thread on the potentiometer shafts could be matched. Before each key cap was screwed onto each shaft, a spring was placed around each shaft. When a key is pressed, the spring compresses, returning the key to the resting position as the user pulls back their finger as shown in Fig. 2.

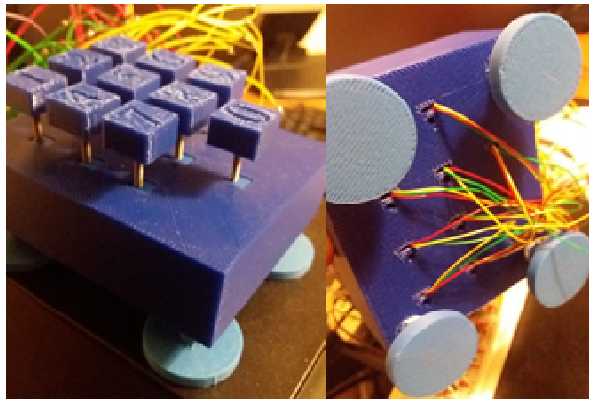

Fig. 2: The top and bottom of the complete number pad. On the left, you can see the potentiometers as the small light blue squares with the shafts protruding up from them. The shafts each have a spring on them, and each key cap is screwed on top. The right picture shows how the wires from the potentiometers lead out through the holes, and the feet.

### B. The Analogue to Digital Converter

This block needs to sample continuously the analogue signal generated by the potentiometer over the course of a button press, at a sample rate sufficiently large to collect a reasonable amount of data. A normal typing speed is 33 words per minute [19], and assuming an average of 4.5 characters in a word, and buttons being pressed around a quarter the time, this gives a button press time of 100ms, with a minimum time of half that. However, because of the size and travel of the buttons used, a button press on the number pad is around five times longer. To record enough data, a minimum of a hundred data values should be gathered on an average button press. Another important consideration is the resolution of the samples. If the resolution is too small or the sample rate too low, differences in button presses will not show up in the data.

The ADCs chosen for this task were the MCP3008 and the MCP3002, being ten bit eight- and two-input analogue to digital converters respectively. Ten bits gives a resolution of just over a thousand values for the full travel of the potentiometers, sufficient for the task. The chips are rated for up to 200k samples per second at 5V supply. These ADCs are controlled via an SPI (serial peripheral interface) bus. A custom protocol was written for this connection because commonly SPI busses deal with 8-bit payloads, not 10-bit.

*C. The comparator block*

This block constantly compares the potentiometer output voltages to a reference voltage, producing a digital signal for each button pressed. The reference voltage is chosen to be sufficiently low to pick up every button press with minimal impact on the resolution (the lowest 10-bit value measured becomes 10 rather than 0 when the digital signal is on, a drop of >1% of the resolution), but also high enough only to activate when the button is pressed. A simple potential divider produced the non-zero reference voltage of around (~0.03V). These signals from the comparator are also fed into the system bus.

*D. The software part*

This part of the design is responsible for user data classification and for user authentication. An off-the shelf-raspberry PI was chosen to implement bot functionalities.

Various algorithms can be used to build a distinctive profile that capture the keystroke characteristics of each user. Possibly the most effective algorithms involve machine learning, which although harder to implement than statistical methods can provide better results [13],[22]. Such methods may require a lot of processing power, possibly too much to be included. Therefore, in this case study we explored two statistics-based approach for data analysis. The first is based on the computation of simple averages and the second is based on probability estimation, with both techniques relying on characteristics extracted from each button press.

With ($S$) corresponding to the displacement value, and ($T$) to the sample number (due to the constant sample rate this is a measure of time), the following characteristics have been extracted:

1) The maximum displacement value ($S_{max}$)

2) The length of button press ($T_{max}$)

3) The plateau length (*PL*)

This is the amount of time for which the displacement is at the maximum value. $T[S_{max0}]$ is the sample number where the displacement hits the maximum value and $T[S_{max1}]$ is when the displacement is no longer maximum. From the results, this tended to be a significant portion of the button press, normally around 50-300 samples. This is calculated based on the three highest resolution levels, to avoid fluctuations in the reading changing the result. When these fluctuations were noticed during testing, they only ever varied between adjacent levels, so having three levels should be enough to remove this risk. It is calculated as follows

$$PL = T[S_{max1}] - T[S_{max0}] \qquad (1)$$

4) The attack value (AV)

This was calculated as the gradient of the press up until the start of the maximum level, so the maximum displacement value divided by the number of samples until the maximum displacement value (within three values, as before) was first reached. It is calculated as follows

$$AV = S_{max0}/T[S_{max0}] \qquad (2)$$

5) The decay value (*DE*)

This was calculated the same way as the attack value but using the number of samples between when the press dropped below the maximum value and the end of the press.

$$DE = S_{max1}/(T_{max} - T[S_{max1}]) \qquad (3)$$

6) The mean average sample (*AV*) is the sum of all displacements, given as follows

$$AV = \Sigma S/T_{max} \qquad (4)$$

The above characteristics can be used to identify/verify the users. The first stage is the enrolment state, wherein training data is collected from each user for a given password (in this case study, 10 repetitions, with the password being 10 digits long).

The second stage is the verification phase, during which a user types their password and an algorithm compares the keystroke profile captured by the hardware part with that associated with the user and stored in the system memory. Such comparisons require the use of a metric. In this case study, we consider two metrics, the first is based on the computation of simple average and the second is based on probability estimation. These two approaches are explained in detailed below.

*a) User Verification using Simple Averages*

In this case, for each of the characteristics outlined above, and for each digit in the password, the mean value is extracted from the training data (ten repetitions of the password), during the enrolment stage. At the following verification phase, when a password is taken for authentication, the differences between the values for the new password and the average values are calculated. This produces a percentage discrepancy for each factor. Logically speaking, if the password was from the same user as the training data, it would be similar to the training data, and so this discrepancy would be low, but if it were another person, the difference would be high. The percentages are then summed to produce a number that is lower if the user is correct, and higher if they are wrong. A threshold can then be set that allows the majority of correct presses through, while rejecting the majority of incorrect ones. This method however does have some flaws, predominantly that there is no prevention against anomalies changing the results. If there is an anomaly in the training data, it will shift the average, and there is the potential for large anomalies with the number pad. If, say, normally the user presses the button about half way, but once in the training data they press it down fully, this will shift the average by 5% for that key. The reverse can also be a problem, that defining features about the wave are averaged out and don't show in the final number, for example if the user always presses to a maximum value of 400 or 600 but not 500, this would be lost and averaged out to 500. However, for a quick comparison between the training data and test data, the averaging method can easily produce a reasonable result and so is used as the first method of comparison. Pseudocode for this method is shown in fig. 3.

```
1  For each characteristic:
2      Calculate characteristic for each key press in each iteration
       of the training data
3      Take the average for each digit over all ten iterations
4      Calculate characteristic for each key press in testing data
5      Calculate percentage difference between the training and
       testing data for each digit
6      Average the percentage differences over the digits to create
       an average percentage difference for the characteristic
7  Sum the percentage differences for each characteristic to produce
   a final value
```

Fig. 3: Pseudocode for the statistical algorithm

*b) User Verification using Probability Estimation*

To overcome the disadvantages of the simple average approach, probability was considered as a second method. It is possible to calculate the probability that a certain value on one of the characteristics was reached by comparing the amount of background data and training data for that value. Background data is data showing how other users press the button, while the training data shows how the correct user presses the button. Each characteristic was then divided into sections that encompassed a range of values. For example, the max level was divided into under 100; 100-199 and so on until over 900 was the tenth category. For each button in the password, and for each characteristic, these categories were then populated with the background data and the training data. The result was two effective bell-curves, one for the background data and one for the user's data. Fig. 4 shows two characteristics and the relevant instances for background and training data. Logically the user's data should vary less assuming the user enters the password in a similar way each time, compared to a background made up of many different people with many different ways of entering the password. This produces a sharper peak on the training data that on the background data.

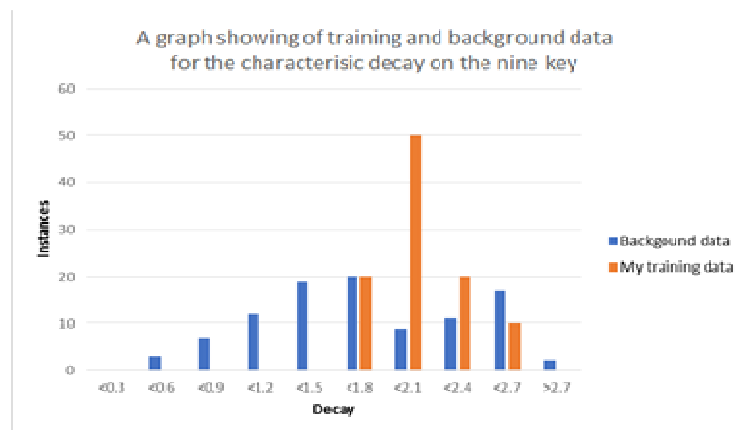

(a)

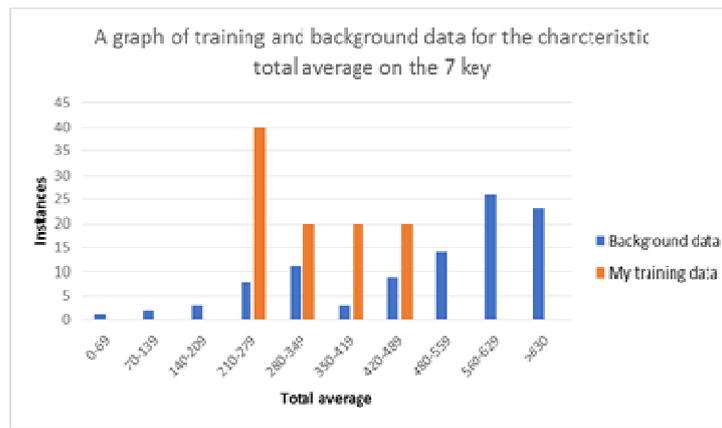

(b)

Fig. 4: Examples of the difference between training data and background data, (a) decay, (b) total average.

To calculate the probability that a characteristic of a press belongs to the user, the background and training data are compared at the value from the press data. First, if there are fewer data sets for the training data than there are for the background, the data must be normalised. In this case, 100 data points were used for background data but only 10 for training data. To weight each instance equally, each instance of the training data is multiplied by the fraction that they differ from the background data by (in this case 10), resulting in the same number of instances for the two types of data. Next, the number of those instances that belong to the user divides the total number of instances in the section that the press data falls into (both user data and background data). This provides a number between zero to one, correlating to the probability of the press belonging to the user rather than to someone else, where zero means that there are no user instances in that section and one means there are no background instances in that section, only user instances. A value of 0.5 means the data is as likely to belong to the user as it is to belong to someone else. A value above 0.5 suggests the data belongs to the user, while a value below suggests the data does not. The relevant pseudocode is shown in fig. 5.

```
1  For each characteristic:
2      Divide each digit into equal possible bands the value could be
3      Count the number of presses from the background data to fall
       into each band
4      Count the number of presses from the training data to fall
       into each band
5      Multiply the training instances by (the amount of background
       data/the amount of training data) so there are an equal number
       of data points in training and background
6      Produce a ratio for each band (no. training instances in that
       band/no. total instances in that band)
7      By looking at which band the testing values fall into,
       calculate an average ratio for that characterisic
8  Average the ratios for all the characteristics to produce a final
   probability value
```

Fig. 5: Pseudocode for the probability based algorithm

## V. EVALUATION

### A. Survey

To collect data to train and test the proposed mechanism, a survey was conducted where people were asked to enter a password onto the number pad. The password was 6193225307 (chosen using https://www.random.org/) and each participant was asked to enter the password onto the number pad in whatever way they wanted. Fourteen people gave eleven repetitions of the password to provide background and testing data, another ten people provided training data and subsequent correct testing data, and eight other people gave data to use as incorrect users. Thirty-two people were involved in the survey, with over 500 repetitions of the password used.

Ten sets of ten passwords were used as the background data for the second algorithm, thus giving 100 instances of each digit in the password against which the probability algorithm could compare. Ten separate users gave training data, and each of them tested the trained algorithms by entering the password another ten times and recording the results. The trained algorithms were then tested by using ten passwords from multiple incorrect users (users whose data was used to test the algorithms were not the same users whose data was used for the background data, and all testing data was unique). In summary, for each user tested, ten passwords trained the algorithm, while a further twenty (ten false and ten correct) tested it.

## B. Results

As expected, the results from the same user were similar to each other (shown in Fig. 2). Two other factors show in the results, firstly that users who pressed the buttons in a specific way (always pressing the button to the maximum, or pressing the buttons quickly, for example) were less distinct from other users doing the same thing compared to others who did not. This is a relatively predictable result, as it shows that the algorithms are identifying characteristics of the presses. The second factor is that the values for correct data varied per user, without necessarily effecting how well that user was recognised. For example, with one user the probability value for correct presses never fell below 0.65 with incorrect presses just above 0.6, whereas with another user correct values fell below 0.45, but incorrect values only reached 0.41. For both of these users, the algorithm completely distinguished the correct and incorrect passwords, but at different parts of the scale.

This shows that no fixed threshold should be used for verification with these algorithms. Therefore, thresholds were chosen for each user by selecting a threshold value such that probabilities greater than the threshold or percentage differences less than the threshold are accepted. The best possible threshold values were chosen, such that the maximum number of correct results was accepted while accepting the minimum number of incorrect results. Error rates for these thresholds could then be calculated by considering the number of falsely accepted or rejected passwords. Table 1. shows these error rates.

TABLE II.  TABLE OF ERROR RATES

| User no. | Probability | | Percentage difference | |
|---|---|---|---|---|
| | *FAR* | *FRR* | *FAR* | *FRR* |
| 1 | 0.00% | 0.00% | 0.00% | 0.00% |
| 2 | 10.00% | 10.00% | 10.00% | 40.00% |
| 3 | 0.00% | 0.00% | 0.00% | 0.00% |
| 4 | 10.00% | 10.00% | 10.00% | 20.00% |
| 5 | 10.00% | 10.00% | 30.00% | 0.00% |
| 6 | 0.00% | 0.00% | 0.00% | 0.00% |
| 7 | 30.00% | 0.00% | 10.00% | 0.00% |
| 8 | 0.00% | 0.00% | 30.00% | 0.00% |
| 9 | 0.00% | 0.00% | 10.00% | 10.00% |
| 10 | 0.00% | 0.00% | 0.00% | 0.00% |
| **Average** | **6.00%** | **3.00%** | **10.00%** | **7.00%** |

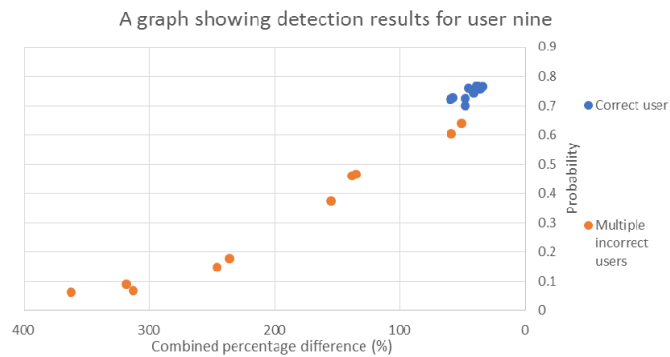

Fig. 6: The results of user nine's training data

Many of the users managed to produce perfect results (four of the ten users for percentage difference, and six of the ten users for probability), whereas some users, for example user 2, produced relatively high error rates. The primary factor for accurate verification was the congruence of button presses. Users who could enter their password the same way every time, both in training and testing produced lower error rates than those whose passwords varied more. This was more of a factor than any particular method of pressing – both users who pressed buttons faster and those who pressed slower were among the results with zero error rates. In addition, worth of note is that the more complex probability algorithm was more successful than the percentage difference algorithm.

An example of the results taken is shown in Fig. 6. The graph shows the clustering of correct data from user nine in the upper right. The incorrect data also clusters into four different groups, each one corresponding to the four different participants who provided incorrect data against which to test. Three of the groups are clearly distinct from the correct data, while the fourth is distinct only in probability. This is likely because user nine and that incorrect user share a similar method of typing, in terms of speed and pressure. The probability algorithm is still able to separate the two, however, whereas the percentage algorithm has some overlap. To calculate errors, ideal thresholds are applied. For probability, this can be >0.65, giving zero errors. For percentage, the best threshold is <58, giving one falsely accepted data point and one falsely rejected data point, so FAR = FRR = 10%.

The average error rates are 4.50% for the probability algorithm and 8.50% for the percentage difference algorithm. By keeping presses consistent, one might expect better results, but it must be mentioned that the thresholds chosen are ideal for the data and therefore may, in practice, give worse results. However, if this system was to be adopted it is expected that a better algorithm could be developed and drastically lower the error rate. For a prototype, these results do suggest very low error rates are possible in practice. Achieving such results with relatively basic algorithms and hardware suggest that a complete system should be successful.

## VI. Conclusion

As previously mentioned, most previous studies into keystroke dynamics only use digital timing data. While this means conventional keyboards or number pads can be used, taking additional data allows for more complex and accurate authentication algorithms, reducing error rates. Studies using static statistical analysis (in the way that this study does) on a keyboard have given error rates from 0.25% to 6% FAR and 1.45% to 12% FRR [13]. Araujo et al. and Bleha et al give examples of such results [16],[27] . These studies used whole keyboards rather than the ten buttons of a number pad, and may have had more complex algorithms for authentication, so the 4.5% average error rate for the probability-based algorithm in this study suggests that the analogue measurements are effective, and a good choice for the limited space available on a hardware token. They are also more effective than the purely maximum pressure-based systems, which only reported 40% average accuracy [25]. Similar previous studies that used neural methods instead of statistical ones showed a notable improvement in the error rates, with a study giving an average error rate of 1% in [28]. This suggests that the proposed system with analogue measurements may be able to achieve even better results using neural methods.

The proposed system has a number of advantages over previous studies. The system can be self-contained - the prototype used a raspberry pi for all computation, and such a processor could be embedded onto the number pad to make it one unit, no bigger than a normal number pad. Studies that have shown great results for analogue data in the past [22], but these have mostly been situated on a computer with an attached analogue keyboard, rather than a self-contained unit. A possible issue with this would be if a large neural net were used for computation, in which case the processing power of an embedded chip may not be sufficient [13, 14]. This could be solved by doing the processing remotely, on the host machine or a server, or by generating the algorithm remotely and then loading the completed algorithm onto the keypad. While the prototype used linear motion potentiometers, a full system could use linear encoders or similar to save on price, and this would be cheaper than the force sensitive resistors of past studies [22, 25].

Overall, this method could increase security by introducing a biometric element to a multi-factor authentication system, relatively cheaply and without being invasive or time consuming for the user. The case study shows how different users can be distinguished by their button presses and gives some simple methods for authenticating people by their button presses. To build on this, applying neural nets or other machine learning techniques to the problem would be a good topic for future research, as the algorithm dictates the effectiveness of the system and machine learning has proved successful in similar problems [13].

Another future direction on the design side include expanding the current implementation to take into consideration temporally close keystrokes, this effectively mean treating the entire password as an n-dimensional feature vector, this may provide better authentication. On the verification side, further analysis may include varying the training-testing ratio and repeating the same analysis above[29, 30]